\documentclass[aps,pra,twocolumn,showpacs]{revtex4-1}
\usepackage{graphicx,amsbsy,amsmath,amsfonts}
\begin{document}
\title{Stationary entanglement in $N$-atom subradiant degenerate
cascade systems}
\author{Massimo Borrelli}
\affiliation{Turku Center for Quantum Physics, Department
of Physics and Astronomy, University of Turku,
20014 Turun Yliopisto, Finland}
\author{Nicola Piovella, Matteo G. A. Paris}
\affiliation{Dipartimento di Fisica, Universit\`a degli Studi di
Milano, I-20133 Milano, Italy}
\begin{abstract}
We address ultracold $N$-atom degenerate cascade systems and show
that stationary subradiant states, already observed in the
semiclassical regime, also exist in a fully quantum regime and for
a small number of atoms. We explicitly evaluate the amount of stationary 
entanglement for
the two-atom configuration and show full inseparability for the
three-atom case. We also show that a continuous variable
description of the systems is not suitable to detect entanglement
due to the nonGaussianity of subradiant states.
\end{abstract}
\date{\today}
\maketitle
\section{Introduction}
Atomic subradiant states have recently gained wide attention
because of their exceptionally slow decoherence
\cite{Dicke,Haroche}. This stability of quantum superpositions
inside the subradiant subspaces originates from the low
probability of photon emission, which means very weak interaction
between the atoms and their environment. Hence, the subradiant
states span a decoherence-free subspace \cite{Zurek,Foldi,Lidar}
of the atomic Hilbert space, and consequently, can become
important from the viewpoint of quantum computation
\cite{Plenio,Beige}.
\par
Subradiance in an extended pencil-shaped sample has been observed
in an unique experiment  in 1985 by Pavolini et al.
\cite{Pavolini}. Among different schemes of multilevel systems,
Crubellier et al.\cite{Cru1,Cru2,Cru3,Cru4} investigated in detail
the arising of subradiance in three-level atoms with a degenerate
transition. More precisely, they considered the case of
three-level atoms with two transitions sharing either the upper
level ('$\Lambda$ configuration) or the lower level ('$V$'
configuration) or the intermediate one ('cascade' configuration),
when both the transitions have the same frequency and the same
polarization. The first two configurations may be experimentally
realized by rather specific atomic level systems, having a small
hyperfine structure, whereas the degenerate cascade configuration,
which contains two cascading transitions of the same frequency,
would in fact be encountered in atomic-level systems only in the
presence of an external field which suitably modifies the atomic
frequencies \cite{Cru2}. Recently, the same degenerate
cascade among three equally spaced levels has been investigated
for collective momentum states of an ultracold atomic gas in a
high-finesse ring cavity driven by a two-frequency laser field
\cite{Piovella}. This system is particularly attractive for the
observation of subradiance, since the absence of Doppler
broadening and collisions at sub-recoil temperatures avoid other
undesirable decoherence mechanisms. Furthermore, the experimental
control of the external parameters, as for instance the
intensities of the pump fields, allows for a continuous tuning
through different subradiant states of the system, which instead
should result difficult or impossible for internal atomic
transitions.
\par
More specifically, the system described in \cite{Piovella}
consists in $N$ two-level ultracold bosonic atoms (at $T\approx
0$) placed in a ring cavity with linewidth $\kappa$ and driven by
two pump fields $E_0$ and $E_1$ of frequency $\omega$ and
$\omega+\Delta$, respectively. The atoms scatter the photons of
the two-frequency pump into a single counterpropagating cavity
mode. The pump frequency $\omega$ is sufficiently detuned from the
atomic resonance $\omega_0$ to neglect absorption. The scattering
process can be described by two steps. In the first step the atom,
initially at rest, scatters the $\omega$ pump photon into a photon
with frequency $\omega_{s1}=\omega-\omega_r$, and recoils with a
momentum $\vec p=\hbar \vec q$ (where $\vec q=\vec k-\vec k_s$ is
the transferred momentum, $\omega_r=\bar q^2/2M$ is the recoil
frequency and $M$ is the atomic mass). In the second step the atom
scatters the $\omega+\Delta$ pump photon into a photon with
frequency $\omega_{s2}=\omega+\Delta-3\omega_r$, changing momentum
from $\vec p=\hbar\vec q$ into ${\vec p}\,'=2\hbar \vec q$. The
recoil shift of $3\omega_r$ arises from  the kinetic energy
conservation, i.e. $\Delta E=(p'^2-p^2)/2M=3\hbar\omega_r$. If the
frequency difference between the two pump fields is
$\Delta=2\omega_r$, then $\omega_{s1}=\omega_{s2}\equiv\omega_s$
and a degenerate cascade between three momentum levels, $\vec
p=m(\hbar \vec q)$, with $m=0,1,2$, is realized. Furthermore, if
the cavity linewidth $\kappa$ is smaller that $2\omega_r$, the
cavity will support only a single frequency $\omega_s$, avoiding
further higher-order scattering with frequency $\omega_s\pm
m\omega_r$. In this way the transitions are restricted to only the
first three momentum states, forming a three-level degenerate
cascade. Although the condition $\kappa\ll 2\omega_r$ is
experimentally demanding, requiring a typical cavity finesse
larger than $10^6$, it is not very far from the state-of-art
optical cavities \cite{Tub,Hammerich}. An important feature of the
momentum three-level cascade is that the ratio of the two
transition rates is $\epsilon=|E_1/E_0|$, i.e. it is an
experimentally tunable parameter, contrarily to the case of atomic
transitions where it is fixed by the branching ratios.
\par
In this paper we address $N$-atom degenerate cascade systems
and show that stationary subradiant states also exist in a fully
quantum regime. Actually, the action of the coupling to the environment 
drives our system into an entangled state without the need of isolating it from
the rest of the universe. We solve explicitly the Master equation governing
the system for a small number of atoms and evaluate the amount of
stationary entanglement \cite{ETC}. 
In turn, our system represents an example of robust generation
of steady state entanglement among atoms via the interaction with
the electromagnetic field which act as an engineered environment 
\cite{2209}.
The paper is structured as follows. In the
next Section we give
a full quantum description of the dynamics and we numerically
solve a quantum master equation for $N=2(3)$-atom systems.
In the Section \ref{s:ent} we study the entanglement properties
of subradiant states looking at the system from both
discrete-variable and continuous-variable (CV) point of view
and show that CV criteria are not able to detect entanglement
due to nonGaussianity of subradiant states. Section \ref{s:out}
closes the paper with some concluding remarks.
\section{Dissipative dynamics and N-atom pure subradiant states}
The three-level degenerate cascade interaction is described by the
following Hamiltonian operator in interaction picture
\begin{equation} H_{int}=-i\hbar g\left[ac_{1}c_{0}^\dagger+\epsilon
ac_{2}c_{1}^\dagger-\textrm{h.c.}\right] \label{hamiltonian}
\end{equation}
for the three atomic bosonic operators, $c_m$, with $m=0,1,2$ and
$[c_m,c^\dagger_n]=\delta_{m,n}$, and for the single cavity
radiation mode, $a$, with $[a,a^\dagger]=1$; $g$ is the coupling
constant and $\epsilon$ is the relative rate of the second
transition, from $m=1$ to $m=2$. The state of the total system is
described by a density operator $\varrho$ which is defined on the
overall Hilbert space $\mathcal{H}=\mathcal{H}_{0}\otimes
\mathcal{H}_{1}\otimes\mathcal{H}_{2}\otimes\mathcal{H}_{f}$,
$\mathcal{H}_{i}$ being the Hilbert-Fock space associated to the
$m=i$ atomic mode and $\mathcal{H}_{f}$ the Hilbert-Fock space
associated to the radiation mode $a$.\\ Introducing the Fock
states $|n_{0},n_{1},n_{2}\rangle$, where $n_{i}$ is the number of
quanta in the $m=i$ mode, it has been demonstrated in
\cite{Cru2,Piovella} that such a $N$-atom system, with $N$ even,
admits $N/2$ subradiant pure states,
\begin{equation}
\begin{aligned}
|sr\rangle_{p}=&C_{p}\sum_{k=0}^{p}\bigg(\frac{-1}
{2\epsilon}\bigg)^{k}\frac{1}{k!}\sqrt{\frac{(2k)!(N-p-k)!}{(p-k)!}}\times\\
&|p-k,2k,N-p-k\rangle
\label{sub_pure}
\end{aligned}
\end{equation}
where
$$
C_p = \left[ \frac{(N-p)!}{p!}\; {_2}F_1(-p\,, 1/2\,;-N + p\,;
\epsilon^{-2}) \right]^{-1/2}
$$
where ${_2}F_1(a,b;c;z)$ is the hypergeometric function and
$p=1,\dots,N/2$, at whom we can add the ground state
$|sr\rangle_0=|0,0,N\rangle$ for $p=0$. If $N\gg1$ these states
exist iff $0\le\epsilon\le1+\sqrt{2}$.
\par
In order to study the time-evolution of $\varrho$ accounting for
scattered photons dissipation too, in a pure quantum mechanical
framework, we introduce the following $T=0$ Lindblad master
equation
\begin{equation} \label{lindblad}
\frac{d}{dt}\varrho=\frac{1}{i\hbar}[H_{int},\varrho]+2\kappa\mathcal{L}
[a] \varrho
\end{equation}
where $\mathcal{L}[a]\varrho=a\varrho
a^{\dagger}-a^{\dagger}a\varrho/2-\varrho a^{\dagger}a/2$ is the
Lindblad operator describing the cavity damping of the mode $a$.
In the next subsections we solve numerically Eq.\eqref{lindblad}
for two and three atoms in terms of total Fock states
$|n_{0},n_{1},n_{2},n\rangle$, where $n$ is the number of
scattered photons. We will find that for every $\epsilon$ and for
$\kappa\ne0$ Eq.\eqref{lindblad} always admits
a stationary
solution of the form
\begin{equation}
\varrho_{s}=\Big[P_{1}|sr\rangle_{11}\langle
sr|+(1-P_1)|sr\rangle_{00}\langle
sr|\Big]\otimes|0\rangle\langle0| \label{rhosub*}
\end{equation}
where $|sr\rangle_{0}, |sr\rangle_{1}$ are the only two populated
states of two-atom and three-atom systems,
$P_1=\textrm{Tr}\left\{\varrho|sr\rangle_{1}\, _{1}\langle
sr|ì\right\}$ and $|0\rangle$ is the vacuum photon state.
\subsection{$N=2$ system}
\begin{figure}[h!]
\includegraphics[width=0.49\columnwidth]{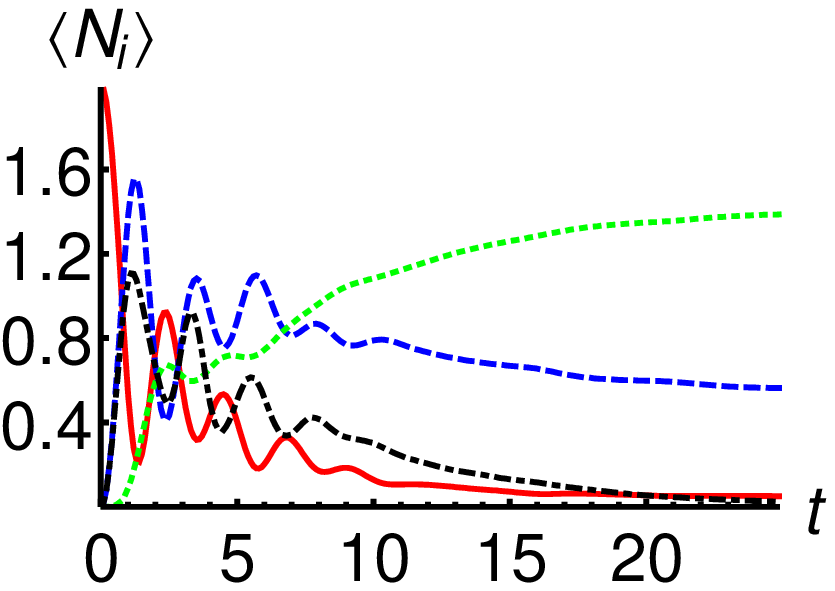}
\includegraphics[width=0.49\columnwidth]{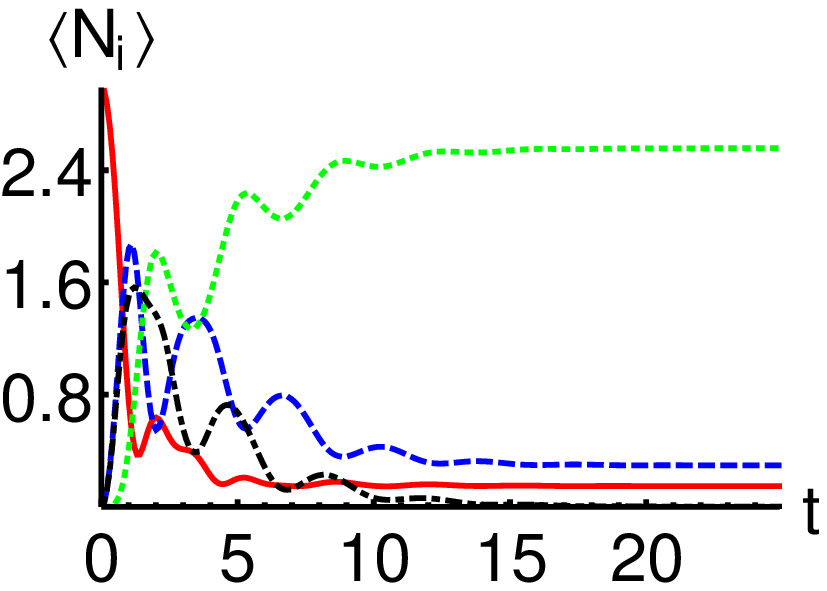}
\caption{Time evolution of $\langle N_{i}\rangle$ for $m=0$ (red),
$m=1$ (blue dashed), $m=2$ (green dotted) and $\langle N_{ph}\rangle$ 
(black dot-dashed) with $t$ in unit of $1/g$. Left: $N=2$, $\kappa=0.2g, 
\epsilon=0.3$. Right: $N=3$, $\kappa=0.3g, \epsilon=0.5$.} \label{avn}
\end{figure}

From Eq.\eqref{sub_pure} it is straightforward to check that the
two-atom system admits the subradiant state
\begin{eqnarray}
|sr\rangle_{1}&=&\frac{1}{\sqrt{1+2\varepsilon^{2}}}(|0,2,0\rangle-\sqrt{2}\varepsilon|1,0,1\rangle).
\label{sub21}
\end{eqnarray}
We choose as initial condition for Eq.\eqref{lindblad}
$\varrho_{i}=|2,0,0,0\rangle\langle2,0,0,0|$. Fig.\ref{avn} (top)
shows the time evolution of the populations $\langle
N_{i}\rangle=\textrm{Tr}[c_{i}^{\dagger}c_{i}\varrho]$ (with
$i=0,1,2$) and $\langle N\rangle=\textrm{Tr}[a^{\dagger}a\varrho]$
for chosen values of $\kappa$ and $\epsilon$: we  observe that
$\langle N_{0}\rangle$ and $\langle N_{1}\rangle$ tend to non-zero
values after the photons have escaped from the cavity.  This
result suggests that the atoms do not decay to the lower momentum
state $m=2$, but they arrange themselves in a superposition of the
three momentum states.  Fig.\ref{prob} (top) shows the time
evolution of the probabilities
$P_{i}=\textrm{Tr}[\varrho|sr\rangle_{ii}\langle sr|]$, for
$i=0,1$, for the same values of $\kappa$ and $\epsilon$.
Asymptotically, $P_{0}+P_{1}=1$ and  $\varrho$ approaches the
steady-state density matrix $\varrho_s$ of Eq.\eqref{rhosub*},
where $P_{1}$ in general is a function of the parameters
$\epsilon$ and $\kappa$. A similar behavior may be observed for
any choice of the interaction parameters. In the bad-cavity limit,
$\kappa\gg g$, Eq. \eqref{lindblad} is well approximated by the
following effective superradiant master equation \cite{Cru2,Boni}
\begin{equation}
\frac{d\varrho_{a}}{dt}=\Gamma S^{-}\varrho_{a}
S^{+}-\frac{\Gamma}{2}(S^{+}S^{-}\varrho_{a}+\varrho_{a}
S^{+}S^{-})\label{ME:SR}
\end{equation}
where $\varrho_{a}$ is the atomic  density operator,
$S^{-}=c_{1}^{\dagger}c_{0}+\varepsilon c_{2}^{\dagger}c_{1}$ and
$\Gamma=g^{2}/ \kappa$.
Solving this equation analytically we get an expression for the
time-evolution of $P_1$ and in the stationary limit we find a
$\Gamma$-independent expression
\begin{equation}
P_1 = \frac{2(1-\epsilon^2)^2}{9\epsilon^{2}+2(1-\epsilon^2)^2}\label{psr}
\end{equation}
This expression is in excellent agreement with the solutions of
Eq. \eqref{lindblad} for any possible value of $\epsilon$ in the
bad cavity regime. In Fig.\ref{ss} (top) we see that $P_{1}$
decreases from unit as $\epsilon$ increases and it vanishes for
$\epsilon=1$, {\rm for which the degenerate three-level cascade
has equal transition rates and the atoms decay superradiantly to
the ground state}. For $\epsilon>1$ the subradiant probability
increases and reaches asymptotically the unity. For large values
of $\epsilon$ the steady-state is no longer a mixed state,
$\varrho_{s}\approx|1,0,1\rangle\langle1,0,1|$.
\subsection{$N=3$ system}
\begin{figure}[h!]
\includegraphics[width=0.49\columnwidth]{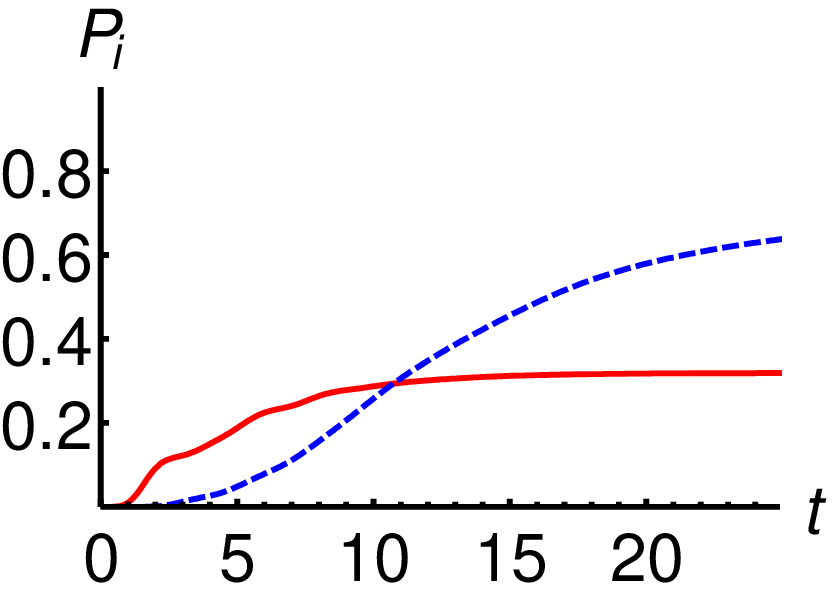}
\includegraphics[width=0.49\columnwidth]{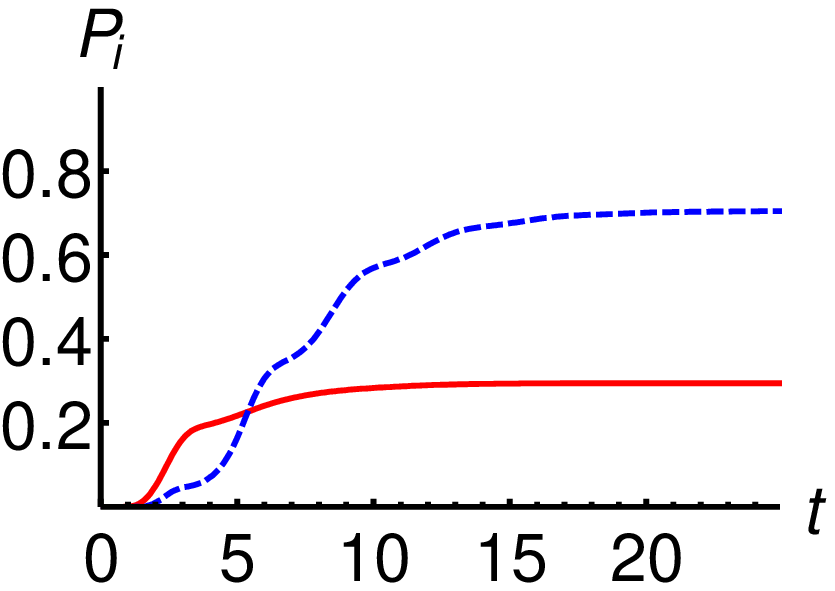}
\caption{Time evolution of $P_{0}$ (blue dotted) and $P_1$
(red) with $t$ in unit of $1/g$. Left: $N=2$, $\kappa=0.2g, 
\epsilon=0.3$; Right: $N=3$, $\kappa=0.3g, \epsilon=0.5$.} \label{prob}
\end{figure}

\begin{figure}[h!]
\includegraphics[width=0.47\columnwidth]{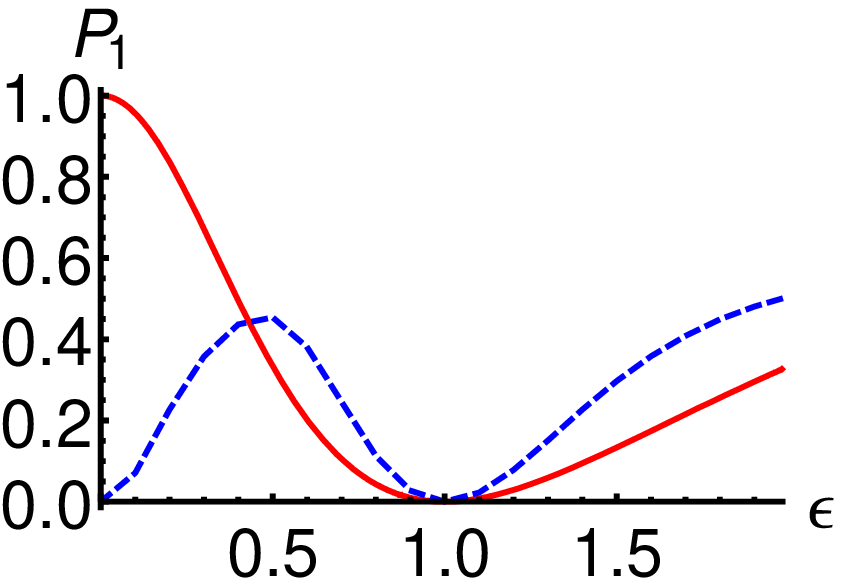}\hfill
\includegraphics[width=0.48\columnwidth]{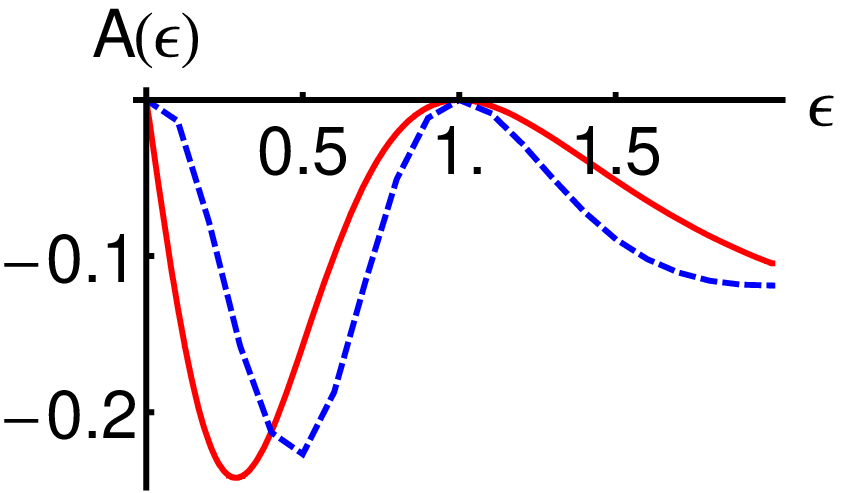}
\caption{Left: steady state probability as a function of $\epsilon$ 
for $N=2$ (red) and $N=3$ (blue dashed). Right: negative eigenvalue of
the partially transposed state as a function of  $\epsilon$ for $N=2$ 
(red) and $N=3$ (blue dashed). In the $N=3$ case we have set  
$\kappa=0.8g$.} \label{ss}
\end{figure}

In this case we cannot directly use the eq.\eqref{sub_pure} which
applies only for $N$ even. The equation $S^{-}|\psi\rangle=0$,
where $|\psi\rangle$ is a generic three-atom state, is satisfied
by the subradiant state
\begin{eqnarray}
|sr\rangle_{1}&=&\frac{1}{\sqrt{1+4\epsilon^{2}}}
(|0,2,1\rangle-2\epsilon|1,0,2\rangle)
\label{sr31}
\end{eqnarray}
Fig.\ref{avn} (bottom) shows the time evolution of the populations $\langle
N_{i}\rangle$ with $i=0,1,2$ and $\langle N\rangle$ for fixed
values of $\epsilon$ and $\kappa$, obtained solving
Eq.\eqref{lindblad} with initial condition
$\varrho_{i}=|3,0,0,0\rangle\langle3,0,0,0|$. As in the $N=2$ case
the mean occupation number of every atomic mode saturates to
non-zero values and, asymptotically, the only atomic states
surviving are the subradiant state $|sr\rangle_1$ and the ground
state $|sr\rangle_{0}=|0,0,3\rangle$. Hence also in this case we
can write the reduced steady-state density operator in the form
(\ref{rhosub*}). Fig.\ref{prob} (bottom) shows $P_1$ vs $t$
 and Fig.\ref{ss} (top) vs $\epsilon$ for a
particular choice of $\kappa$. There are some important
differences and analogies with the $N=2$ case. First we note that
this time $P_1$ vanishes for $\epsilon=0$, whereas in the $N=2$
case it has a maximum and is equal to unity. This is a consequence
of the fact that the two-atom subradiant state (\ref{sub21})
contains the state $|0,2,0\rangle$ which is the ground state for
$\epsilon=0$ (only the $m=0\rightarrow m=1$ transition can take
place). On the contrary, for $N=3$ the steady-state density
operator does not contain the state $|0,3,0\rangle$ which is the
three-atom analogue; moreover, it is a linear superposition of
states with $n_{2}\ne 0$ so, if the second transition is
forbidden, it is impossible to populate these states observing
subradiance. Secondly, whereas for $N=2$ $P_1$ decreases
increasing $\epsilon$ and vanishes for $\epsilon=1$, for $N=3$
$P_1$ increases, it reaches a maximum around $\epsilon\approx 0.5$
and decreases again falling off to zero for $\epsilon=1$. For
$\epsilon>1$ the behavior of the two plots is quite similar.
\subsection{Large $N$ and p-degeneracy}
Let us consider now the $N\gg1$. In \cite{Piovella} the following
relation linking the $p$-index and $\epsilon$ has been given:
\begin{align}
p & =\frac{N}{2}\left(\frac{1-2\epsilon^2}{1-\epsilon^2}\right)
\qquad
0\leq\epsilon\leq 1/\sqrt{3} \label{pe1} \\
p & =N\left(\frac{1-\epsilon^2}{1+\epsilon^2}\right)^2 \quad
1/\sqrt{3}\leq\epsilon\leq 1+ \sqrt{2}
\label{pe2}
\end{align}
Let us pay more attention to formulas \eqref{pe2}. For
$1/\sqrt{3}\leq\epsilon\leq\sqrt{3}$ it is easy to prove that they
exist two different values of $\epsilon$ corresponding to the same
value of $p$ \rm{(except for $\epsilon=1$ yielding $p=0$ i.e.
the ground state)}. This means that, once $p$ is fixed, the two
subradiant states $|sr\rangle_{p(\epsilon_{0})},
|sr\rangle_{p(\epsilon_{1})}$ are both acceptable,
provided that they are actually different states.
Now, let us assume that, for a fixed $p$, \rm{it is possible
to build} the following two states:
\begin{equation}
|\psi^{(\pm)}_{p}\rangle=\frac{1}{\sqrt{2}}(|sr\rangle_{p(\epsilon_{0})}\pm|sr\rangle_{p(\epsilon_{1})}).
\label{qub1}
\end{equation}
It is straightforward to show that the two following states are orthonormal
\begin{equation}
|\phi^{(\pm)}_{p}\rangle=\frac{1}{\sqrt{1\pm\alpha}}|\psi^{(\pm)}_{p}\rangle
\label{qub2}
\end{equation}
where $\alpha=\,_{p(\epsilon_{0})}\langle
sr|sr\rangle_{p(\epsilon_{1})} =\,_{p(\epsilon_{1})}\langle
sr|sr\rangle_{p(\epsilon_{0})}\in\mathbb{R}$. We can compute the
total kinetic energy of the condensate and also the energy
difference between these two states
$|\Delta_{p}|=|E_{p}^{(+)}(\epsilon)-E_{p}^{(-)}(\epsilon)|$ which
turns out to be
\begin{equation}
\begin{aligned}
|\Delta_{p}|=&\bigg|\frac{4(2N-2p-2[\bar{k}(\epsilon_{0}(p))+\bar{k}(\epsilon_{1}(p))])}{\sqrt{2(1+\alpha)}}-\\
&\frac{2[\bar{k}(\epsilon_{0}(p))+\bar{k}(\epsilon_{1}(p))]}{\sqrt{2(1-\alpha)}}\bigg|
\end{aligned}
\end{equation}
where
$$
\begin{aligned}
\bar{k}(\epsilon_{j}(p)) =&\frac{p! (N-p-1)!}{2 \epsilon_{j}^2 (N-p)! (p-1)!}\times\\
&\frac{{_2}F_1(1-p,3/2; 1+p-N; \epsilon_{j}^{-2})} {{_2}F_1(-p,
1/2;p-N; \epsilon_{j}^{-2})}.
\end{aligned}
$$
Since this parameter depends on $\epsilon$, in principle it should
be possible to distinguish between the two states \eqref{qub2}.
More interesting, by applying an external potential
coupling these two states, we could realize a fictitious two-level
system where the many-body states $|\phi^{(\pm)}_{p}\rangle$ play
the role of the logical-basis states $|0\rangle, |1\rangle$. A
similar system has been proposed in \cite{Keitel} for a 
$\Lambda$ configuration.
\section{Entanglement properties}\label{s:ent}
In this section we address the entanglement properties of the
 state \eqref{rhosub*} either regarded as made of three
bosonic modes or, upon considering the selection rule induced by
the constraint on the number of atoms, as a finite dimensional
one. Generally speaking, there are several classes of entanglement
in a tripartite quantum system, from genuine tripartite
entanglement to full separability passing through bipartite
separability, which are defined on the basis of all  possible
groupings of the three parties \cite{tre}. Our primary goal is to
establish whether the stationary state is fully inseparable and if
it is possible to quantify the amount of entanglement.
\par
At first, we notice that our system consists of three atomic
modes for which, being the atomic ensemble made of a finite and
fixed number of atoms, only a finite number of quanta per mode is
permitted from a minimum of $n_{i}=0$ up to a maximum of
$n_{i}=N$. This physical constraint corresponds to a geometrical
constraint which identifies a subspace $\mathcal{S}$ of
$\mathcal{H}_{a}=\mathcal{H}_{0}\otimes
\mathcal{H}_{1}\otimes\mathcal{H}_{2}$ defined as
$$\mathcal{S}=\left\{\varrho\in\mathcal{H}\ :
\sum_{i=0}^{2}\langle N_{i}\rangle=N\right\}\:.$$ Hence, being
$\mathcal{S}$ a finite dimension subspace, we can make use of the
PPT (\textit{Positive Partial Transposition}) criterion for
discrete-variable quantum systems. In fact, PPT condition is
sufficient to establish entanglement of a state and, as we will
see, it is enough to show that $\varrho_s$ is genuinely tripartite
entangled. Partially transposing  with respect to a given party
corresponds to a specific grouping of two parties with respect to
the third: for example when we partial-transpose with respect to
the $m=0$ mode we are grouping together the $m=1$ and $m=2$ modes
with respect to $m=0$. Suppose we have partial-transposed with
respect to $m=0$, obtaining $\varrho_{s}^{T_0}$ and let the
eigenvalues of $\varrho_{s}^{T_0}$ be $\{\lambda_{k}^{(0)}\}$. If
it exists at least one $k$ such that $\lambda_{k}<0$ we can
conclude that
$\varrho_{s}\ne\sum_{j}p_{j}\,\varrho_{j}^{0}\otimes\varrho_{j}^{12}$.
If the positivity condition is violated for $\varrho_{s}^{T_{1}}$
and $\varrho_{s}^{T_{2}}$ too we can state  that $\varrho_s$ is
genuinely tripartite entangled. We start with the $N=2$ case and
we take the reduced density operator
$\textrm{Tr}_{ph}[\varrho_{s}]$ where $\varrho_{s}$ is given by
eq.\eqref{rhosub*} with $P_1$ given by eq.\eqref{psr}. The
$\mathcal{S}$ subspace is spanned by six Fock states:
$|2,0,0\rangle, |0,2,0\rangle, |0,0,2\rangle, |1,1,0\rangle,
|1,0,1\rangle, |0,1,1\rangle$. If we compute the eigenvalues of
every $\varrho_{s}^{T_{j}}$ with $j=0,1,2$ we find  that it always
exists a negative eigenvalue $A_2(\epsilon)$
\begin{align}
A_2(\epsilon) & =-\frac{\sqrt{2}}{2}
\frac{\epsilon(\epsilon^{2}-1)^{2}}{(\frac{1}{2}+
\epsilon^{2})^{2}(2+\epsilon^{2})} \notag \\ &=
-\frac{\epsilon^{2}(2\epsilon^{2}+3)^{2}+2}{4\sqrt{2}
(\epsilon^{2}-1)^{2}}\langle N_{0}\rangle\langle N_{1}\rangle
\label{autoval1}
\end{align}
where the expression on the second line provides a link between the negative
eigenvalue, which directly quantifies the amount of entanglement, and a
couple of moments which are, at least in principle, amenable to direct
measurements.
\par
In the $N=3$ case $\mathcal{S}$ is spanned by the following ten
states: $|3,0,0\rangle$, $|0,3,0\rangle$, $|0,0,3\rangle$,
$|2,1,0\rangle$, $|2,0,1\rangle$, $|0,1,2\rangle$,
$|0,2,1\rangle$, $|1,0,2\rangle$, $|1,2,0\rangle$, $|1,1,1\rangle$
and the PPT condition is still quite simple to employ to establish
tripartite entanglement. On the other hand, for $N=3$ we have not
been able to give an analytical expression for $P_1$ and therefore
we resort numerical evaluation of the eigenvalues. As in the
previous case we find a negative eigenvalue $A_3(\epsilon)$ which
turns out to be always the same for every possible partial
transposition and for any value of $\epsilon$. This means that
also the three-atom $\varrho_{s}$ is genuinely tripartite
entangled. Fig.\ref{ss} (bottom) shows the behavior of $A_N(\epsilon)$
both for $N=2$ and $N=3$. The behavior is qualitatively the same
for both cases. The eigenvalues decrease from zero and reach a
minimum after which they start to increase. As we should expect
they vanish for $\epsilon=1$ and after this they start to decrease
again.
\begin{figure}[h!]
\centering
\includegraphics[width=0.48\columnwidth]{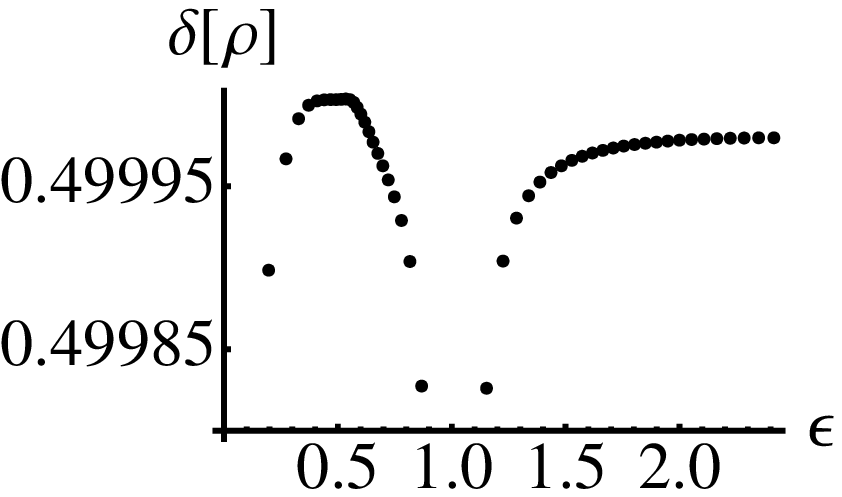}
\includegraphics[width=0.48\columnwidth]{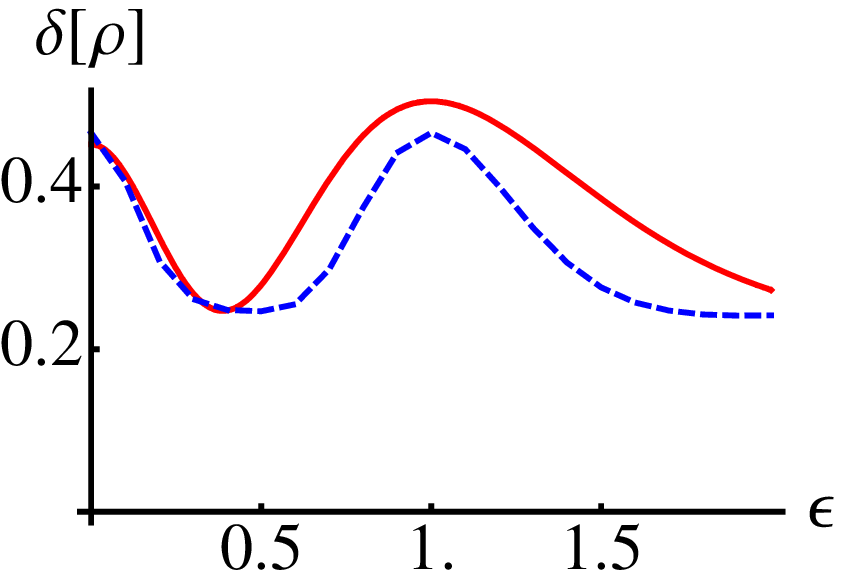}
\caption{Up: $\delta[\varrho]$ for $N=50$ vs. $\epsilon$ accounting for relations \eqref{pe1}-\eqref{pe2}. Bottom: $\delta[\varrho]$ for $N=2$ (red) and $N=3$ (blue dashed); in this second case we have set  $\kappa=0.8g$.} \label{nong}
\end{figure}
\par
\subsection{Continuous variable description}
Let us now investigate the entanglement properties of our system
as a continuous variable one \cite{Wal09}, i.e. taking into account that we are
dealing with three bosonic modes describing the number of atoms
within a given state (for instance momentum state). To this aim we
start by reviewing few facts about the description of a multimode
continuous variable system \cite{GSI}. Upon introducing the
canonical operators 
$q_{j}=(c_{j}+c_{j}^{\dagger})/\sqrt{2}$ and
$p_{j}=i(c_{j}^{\dagger}-c_j)/\sqrt{2}$ in terms of the
single-mode operators $c_j$, $j=0,1,2$ we may write the vector of
operators $\boldsymbol{R}=(q_{1},p_{1},q_2,p_2,q_3,p_3)^{T}$. The
so-called vector of mean values
$\boldsymbol{X}=\boldsymbol{X}[\varrho]$ and covariance matrix (CM)
$\boldsymbol{\sigma}=\boldsymbol{\sigma}[\varrho]$ of a quantum
state $\varrho$ have components $X_{j}=\langle R_{j}\rangle$ and
$R\sigma_{jk}=\frac{1}{2}\langle \{ R_{j},R_{k}\}\rangle-\langle
R_{j}\rangle\langle R_{j}\rangle$ where  $\langle
O\rangle=\textrm{Tr}[\varrho\, O]$ and $\{A,B\}=AB+BA$ denotes the
anticommutator. The characteristic function
$\chi[\varrho](\boldsymbol{\lambda})=\textrm{Tr}[\varrho\,
D(\boldsymbol{\lambda})]$ is defined in terms of the multimode
displacement operator
$D(\boldsymbol{\lambda})=\otimes_{j=1}^{n}D(\lambda_{j})$, with
$\boldsymbol{\lambda} =(\lambda_{1},\cdots,\lambda_{n})^{T}$,
$\lambda_{j}\in\mathbb{C}$ and
$D(\lambda_{j})=\exp\{\lambda_{j}c_{j}^{\dagger}-\lambda_{j}^{*}c_{j}\}$.
A quantum state $\varrho_{G}$ is referred to as a Gaussian state
if its characteristic function has the Gaussian form
$\chi[\varrho_{G}](\boldsymbol{\Lambda})=\exp\left\{-\frac{1}{2}
\boldsymbol{\Lambda}^{T}\boldsymbol{\sigma}\boldsymbol{\Lambda}
+\boldsymbol{X}^{T}\boldsymbol{\Omega}\boldsymbol{\Lambda}\right\}$
where $\boldsymbol{\Lambda}$ is the real vector
$\boldsymbol{\Lambda}
=(\mathfrak{Re}\lambda_{1},\mathfrak{Im}\lambda_{1},\cdots,\mathfrak{Re}
\lambda_{n},\mathfrak{Im}\lambda_{n})$ and $\boldsymbol{\Omega}$
is the symplectic matrix
$\boldsymbol{\Omega}=i\oplus_{j=1}^{n}\sigma_{2}$, $\sigma_{2}$
being the $y$ Pauli matrix. Gaussian states are thus completely
characterized by their mean values vector and their covariance
matrix. For bipartite Gaussian states, the PPT condition is
necessary and sufficient for separability \cite{Sim}, and it may
be rewritten in terms of the CM, which uniquely
determines the entanglement properties of the state under
investigation. A mode of a Gaussian state is separable from the
others iff $\Lambda_j\boldsymbol{\sigma}\Lambda_j + \frac{i}{2}
\boldsymbol{\Omega} \geq 0$, where $\Lambda_j$ is a diagonal
matrix implementing the partial transposition at the level of
CM, i.e. inverting the sign of the $j$-th momentum.
In our case $\Lambda_1=\hbox{Diag}(1,-1,1,1,1,1)$,
$\Lambda_2=\hbox{Diag}(1,1,1,-1,1,1)$,
$\Lambda_3=\hbox{Diag}(1,1,1,1,1,-1)$. When we have a non Gaussian
state the violation of the above condition is still a sufficient
condition for entanglement, whereas the necessary part is lost.
\par
Let us start by focusing attention on a $N$-atom pure subradiant
state (\ref{sub_pure}). The density operator is
$\varrho_{p}=|sr\rangle_{pp}\langle sr|$ and the only
non-vanishing terms of the CM
$\boldsymbol{\sigma}[\varrho_{p}]$ are the diagonal ones, 
which are reported in appendix \ref{a:cm}.
This fact is a direct consequence of two selection rules affecting
the geometrical form of subradiant states: the first one is the
conservation of the number of atoms and the second one is the
presence of the $k$-index in every Fock component of
$|sr\rangle_{p}$. As a matter of fact, the CM
$\boldsymbol{\sigma} [\varrho_{p}]$ satisfies the three
separability conditions $\Lambda_j \boldsymbol{\sigma}\Lambda_j +
\frac{i}{2} \boldsymbol{\Omega} \geq 0$, $j=1,2,3$ and therefore
we have no information about the entanglement properties of
$\varrho_p$, i.e. we cannot use the PPT condition translated to
continuous-variable systems to analyze and quantify entanglement.
On the other hand, we know from the previous analysis that
$\varrho_p$ is genuinely tripartite entangled state and this
suggests that it should be far from being a Gaussian state.
Indeed, the characteristic function is given by a Gaussian modulated 
by Laguerre polynomials  and thus 
shows a distinctive nonGaussian shape. 
\par
In the following we will evaluate quantitatively the nonGaussianity 
of the subradiant states upon the use of the nonGaussianity (nonG) 
measure introduced in \cite{NG1}, which is defined as
\begin{equation}
\delta[\varrho]=\frac{D_{HS}^{2}[\varrho,\tau]}{\mu_\varrho}
=\frac{\mu_\varrho+\mu_\tau-2\kappa_{\varrho\tau}}{2\mu_\varrho}
\label{nonG}
\end{equation}
where $D_{HS}[\varrho,\tau]$ denote Hilbert-Schmidt distance
between the state $\varrho$ under scrutiny and its
reference Gaussian state $\tau$, i.e. a Gaussian state  with the same
mean value $\boldsymbol{X}[\tau]= \boldsymbol{X}[\varrho]$ and
CM $\boldsymbol{\sigma}[\tau]=
\boldsymbol{\sigma}[\varrho]$, 
$\mu_\varrho=\textrm{Tr}[\varrho^{2}]$ is the purity of
$\varrho$, and $\kappa_{\varrho\tau}=\textrm{Tr}[\varrho\,\tau]$
is the overlap between $\varrho$ and $\tau$. The nonG measure
$\delta[\varrho]$ possesses all the properties for a good measure
of the non-Gaussian character and, in particular,
$\delta[\varrho]=0$ iff $\varrho$ is a Gaussian state. For the
pure subradiant states $\varrho_{p}=|sr\rangle_{pp}\langle sr|$
the reference Gaussian state has the form
$\tau=\nu_{0}\otimes\nu_{1}\otimes\nu_{2}$ where
$\nu_{j}=(1-y_j)\sum_s y_j^s |s\rangle\langle s|$,
$y_j=N_j/(1+N_j)$ is a (Gaussian) thermal state with $N_j$ average
thermal photons. For a three-mode thermal state we have
$\sigma_{11}=\sigma_{22}=\frac12 (N_0+\frac12)$,
$\sigma_{33}=\sigma_{44}=\frac12 (N_1+\frac12)$,
$\sigma_{55}=\sigma_{66}=\frac12 (N_2+\frac12)$. Since for a
subradiant state the average occupation numbers are given by
$N_{0}=p-\langle k\rangle_{p}$, $N_{1}=2\langle k\rangle_{p}$,
$N_{2}=N-p-\langle k\rangle_{p}$ we have, after some calculations,
that the non-Gaussianity of $\varrho_{p}$ is given by
\begin{align}
\delta[\varrho_{p}]&=\frac{1}{2}+\frac{1}{2}\prod_{j=0}^{2}
\frac{1-y_j}{1+y_{j}}-|C_{p}|^2 \prod_{j=0}^2 (1-y_j)
\notag \\&\times\sum_{k=0}^{p}\beta_{k}^{2}\, y_0^{p-k}\,y_1^{2k}\, y_2^{N-p-k}
\end{align}
For a large number of atoms $N\gg 1$ the nonGaussianity of
$\delta[\varrho_{p}]$ may be evaluated upon exploiting the
Eq. \eqref{pe1}\eqref{pe2} i.e. the relation between $p$ and $\epsilon$.
In practice, every value of $\epsilon$ corresponds to a single $p$
and in turn to the pure subradiant state $\varrho_p$. 
In Fig.\ref{nong} (top) we show the behavior of the nonGaussianity
$\delta[\varrho_{p}]$ for $N=50$ and
$0\leq\epsilon\leq1+\sqrt{2}$. As it is apparent from the plot,
the subradiant states are always non-Gaussian and the value of
$\delta$ is almost constant. For $N=2,3$ it is possible to
evaluate explicitly the non-gaussianity for the stationary state
\eqref{rhosub*}. Since the number of atoms is small $p$ and
$\epsilon$ are independent one another and we are not allowed to
use relations \eqref{pe1}-\eqref{pe2}. For $N=2,3$ we have the ground state,
$\varrho_0$, and a single subradiant state, $\varrho_1$, and the
nonGaussianity may be expressed as
\begin{align}
\delta[\varrho_{s}]=P_{1}(P_{1}-1)+\frac{1+\mu_\tau}{2}-(1
-P_{1})\kappa_{\varrho_{0}\tau}
-P_{1}\kappa_{\varrho_{1}\tau}\notag\:.
\end{align}
The reference Gaussian state is again a three-mode thermal state, as it can
be easily demonstrated by noticing that the CM 
of $\varrho_s$ is the convex combination of those
corresponding to $\varrho_0$ and $\varrho_1$ i.e.
$\boldsymbol{\sigma}[\varrho_s]=P_1 \boldsymbol{\sigma}[\varrho_1] +
(1-P_1)\boldsymbol{\sigma}[\varrho_0]=P_1$.
$P_{1}$ is given by eq.\eqref{psr} for $N=2$ and it may be
evaluated numerically for $N=3$. The overlaps are given in 
appendix \ref{a:ov}.
The behaviour of the nonGaussianity 
$\delta[\varrho_{s}]$ as a function of $\epsilon$ for $N=2,3$ 
is shown in Fig. \ref{nong} (bottom). 
Notice that the behaviour of $\delta[\varrho_{s}]$ does not show 
a strict correlation to that of the negative eigenvalue $A(\epsilon)$ 
and thus we cannot use nonGaussianity to assess quantitatively 
the failure of CV PPT condition in detecting stationary entanglement in
our system.
\section{Conclusions} \label{s:out}
In this paper we have considered a systems made of $N$ two-level
ultracold bosonic atoms in a ring cavity where subradiance appears in
the bad cavity limit. We provided a full quantum description of the
dynamics and have shown the appearance of stationary entanglement among
atoms.  We have evaluated the amount of steady state entanglement for $N=2$ and $N=3$
atoms as quantified by the negativity of the partially transposed
density matrix. We have also investigated the entanglement properties of
the systems as a continuous variable one and have shown that it is not
possible to detect entanglement due to the nonGaussian character of
subradiant states.  It is important to notice that the life time of
experimentally demonstrated entangled states is generally limited, due
to their fragility under decoherence and dissipation. Therefore, in
order to decrease dissipation and decoherence, strict isolation from the
environment is usually considered. On the contrary, in our system, the
action of the coupling to the environment drives it into a steady
entangled state, which is robust with respect to the interaction
parameters. Our results pave the way for entanglement characterization
\cite{entch} of subradiant states and suggest further investigations for 
large number of atoms.
\section*{Acknowledgments}
The authors thanks M. M. Cola for useful discussions.  MGAP gratefully 
acknowledges support from the Finnish cultural foundation.
\appendix
\section{CM of the subradiant state}
\label{a:cm}
The covariance matrix of the subradiant state $\varrho_{p}=|sr
\rangle_{pp}\langle sr|$ is diagonal. The nonzero matrix elements
may be expressed as
\begin{align}
\sigma_{11}&=\sigma_{22}=\frac{|C_{p}|^{2}}{2}
\sum_{k=0}^{p}\beta_{k}^{2}\left(p-k+\frac{1}{2}\right)
\\ &=
\frac{1}{2}\left(p-\langle k\rangle_{p}-\frac{1}{2}\right)
\notag\\
\sigma_{33}&=\sigma_{44}=\frac{|C_{p}|^{2}}{4}
\sum_{k=0}^{p}\beta_{k}^{2}\left(4k+1\right)
\\ & = \langle k\rangle_{p}+\frac{1}{4}
\notag\\
\sigma_{55}&=\sigma_{66}=\frac{|C_{p}|^{2}}{2}
\sum_{k=0}^{p}\beta_{k}^{2}\left(N-p-k+\frac{1}{2}\right)
\\ & = 
\frac{1}{2}\left(N-p-\langle k\rangle_{p}+\frac{1}{2}\right)
\notag
\end{align}
where we have defined
$$
\beta_{k}= \left(-\frac{1}{2\epsilon}\right)^{k}
\left(\frac{1}{k!}\right)\sqrt{\frac{(2k)!(N-p-k)!}{(p-k)!}}
$$
and $\langle k\rangle_{p}=|C_{p}|^{2}\sum_{k=0}^{p}\beta_{k}^{2}k$
is the mean value of $k$ for the $p$-subradiant state.
\section{Overlaps}
\label{a:ov}
The overlaps $\kappa_{\varrho_j\tau}^{(N)}$ between the states 
$\varrho_j$ and their reference Gaussian states for $N$ atoms, 
which appears in the expression of the nonGaussianity 
$\delta[\varrho_{s}]$ are given by
\begin{align}
\kappa^{(2)}_{\varrho_{0},\tau}&=
\kappa^{(3)}_{\varrho_{0},\tau}=
y_2^2 \, \prod_{j=0}^{2} \frac{1-y_j}{1+y_j}
\notag\\
\kappa^{(2)}_{\varrho_{1},\tau}&=\frac{
y_1^2+2\epsilon^{2}y_0\, y_2 }{1+2\epsilon^{2}}
\prod_{j=0}^{2} \frac{1-y_j}{1+y_j}
\notag \\
\kappa^{(3)}_{\varrho_{1},\tau}
&=\frac{y_2
\left(y_1^2+4\epsilon^{2}y_0\, y_2\right)}{1+4\epsilon^{2}}
\prod_{j=0}^{2} \frac{1-y_j}{1+y_j}\:.
\end{align}

\end{document}